\documentclass{sf2a-conf2014}
\usepackage{graphicx}
\usepackage{hyperref}
\usepackage[]{natbib}  
\usepackage{epstopdf}
\usepackage{color}

\def\BibTeX{{\rm B\kern-.05em{\sc i\kern-.025em b}\kern-.08em
    T\kern-.1667em\lower.7ex\hbox{E}\kern-.125emX}}
\bibpunct{(}{)}{;}{a}{}{,}  

\begin{document}

\TitreGlobal{SF2A 2014}


\title{Using the Virtual Observatory: multi-instrument, multi-wavelength study of high-energy sources}

\runningtitle{Using the VO to explore high energy sources}

\author{S.~Derri\`ere$^1$}
\author{R.~W.~Goosmann$^1$}
\author{C.~Bot$^1$}
\author{F.~Bonnarel}\address{Observatoire Astronomique de Strasbourg, Universit\'e de Strasbourg, CNRS, UMR 7550, 11 rue de l'Universit\'e, 67000 Strasbourg, France}

\setcounter{page}{237}


\maketitle


\begin{abstract}
This paper presents a tutorial explaining the use of Virtual Observatory tools in high energy astrophysics. Most of the tools used in this paper 
were developed at the Strasbourg astronomical Data Center and we show how they can be applied to conduct a multi-instrument, multi-wavelength analysis of sources detected by the High Energy Stereoscopic System and the Fermi Large Area Telescope. The analysis involves queries of different data catalogs, selection and cross-correlation techniques on multi-waveband images, and the construction of high energy color-color plots and multi-wavelength spectra. The tutorial is publicly available on the website of the European Virtual Observatory project\footnote{http://www.euro-vo.org/?q=science/scientific-tutorials}.
\end{abstract}



\section{Introduction}

The Virtual Observatory (VO) offers a vast set of tools\footnote{http://www.ivoa.net/astronomers/applications.html} to explore archival observational data at almost all observable wavelengths. In this note, we illustrate the capacities of the VO in the gamma ray domain. We provide a step-by-step tutorial to walk the reader through the application of data mining tools as well as cross-matching procedures involving observational data from other wavebands. The user may explore how to...

\begin{itemize}
\item ...query astronomical catalogs in different gamma-ray bands using VO tools
\item ...cross-correlate catalogs to find an object at different photon energy bands
\item ...apply selection criteria when extracting sources from a catalog
\item ...use the observational measures of the selected objects to explore possible correlations
\item ...visualize astronomical images from the radio up to the high energy domain
\item ...display spectral energy distributions obtained from different photometric data sets
\end{itemize}

\section{Source catalogs in the H.E.S.S. and the Fermi-LAT energy bands}

\subsection{Exploring the gamma ray sky from the ground and from space}

The {\it High Energy Stereoscopic System (H.E.S.S.)} is a ground-based gamma-ray observatory originally consisting of four optical Cherenkov telescopes. They are situated on top of the Gamsberg mountain in Namibia\footnote{http://www.mpi-hd.mpg.de/hfm/HESS} and detect air showers triggered by gamma ray photons in the energy range of 10s of GeV to 10s of TeV. The latest upgrade of the observatory came in summer 2012 when a fifth, larger telescope was added to constitute H.E.S.S.~II.

The {\it Fermi Gamma-ray Space Telescope} is a space-borne X-ray/gamma-ray observatory detecting sources in an energy range of 8~keV up to more than 300~GeV\footnote{http://fermi.gsfc.nasa.gov}. It carries two instruments, the {\it Glast burst monitor} reporting gamma ray bursts and the {\it Fermi Large Area Telescope (Fermi-LAT)} observing the transient gamma ray sky across a large field of view. The satellite was launched in 2008.

Both observatories are complementary in studying the non-thermal universe. Thermal heating processes can produce at maximum X-ray radiation up to a few keV. Beyond these photon energies, the radiation must be produced by nuclear processes or accelerated particles, for instance in shocks or magnetic reconnection events. The list of objects radiating in the gamma ray sky includes supernovae, pulsars, astrophysical jets in active galactic nuclei or X-ray binaries, gamma ray bursts, and our sun.

\subsection{H.E.S.S., Fermi-LAT and the Virtual Observatory}

For both {\it H.E.S.S.} and {\it Fermi-LAT}, public catalogs of observed sources have been compiled \citep{Carrigan2013,Gasparrini2012} and can be queried with the help of VO tools. It is the ultimate goal of the VO to grant access to the whole body of astronomical observations at all wavebands. Therefore, a unifying and versatile data format was developed, the {\it VOTable} format \citep{Ochsenbein2011}. {\it VOTable} is an XML-based standard allowing to embed both tabular data and the corresponding metadata (e.g. column description) in a single file. Most VO-compliant data tools are able to understand and provide outputs in {\it VOTable} format. This way, the interoperability between different tools and wavebands can be assured. 

The step-by-step tutorial summarized in the following sections makes great use of the {\it VOTable} format. We illustrate only the main steps, of which a complete and more detailed version can be found on the web\footnote{http://www.euro-vo.org/?q=science/scientific-tutorials}.

\section{Cross-matching H.E.S.S. and Fermi-LAT to identify multi-wavelength sources}

\begin{figure*}
  \centering
  \includegraphics[width=9.5cm]{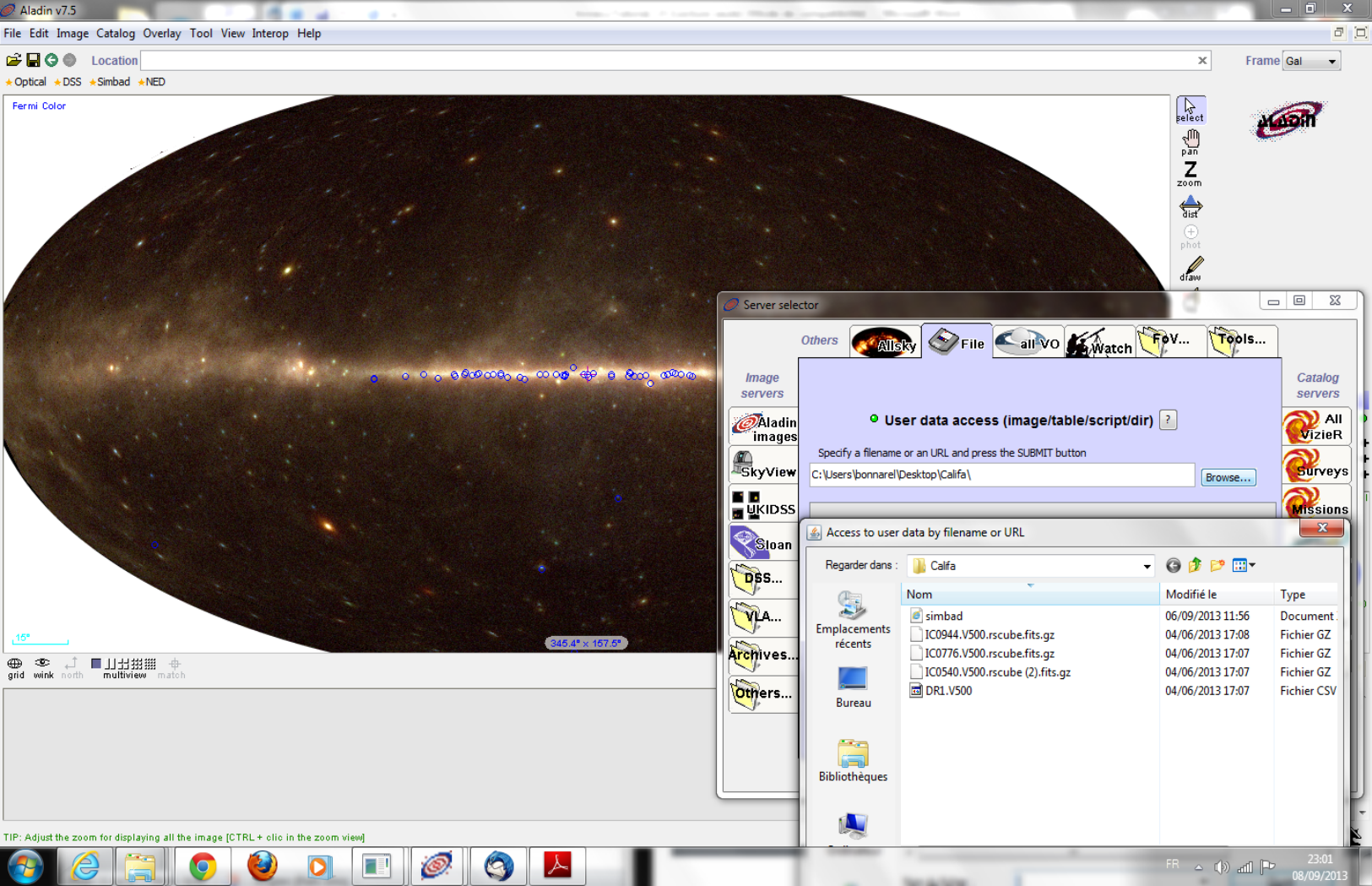}
  \includegraphics[width=6.5cm]{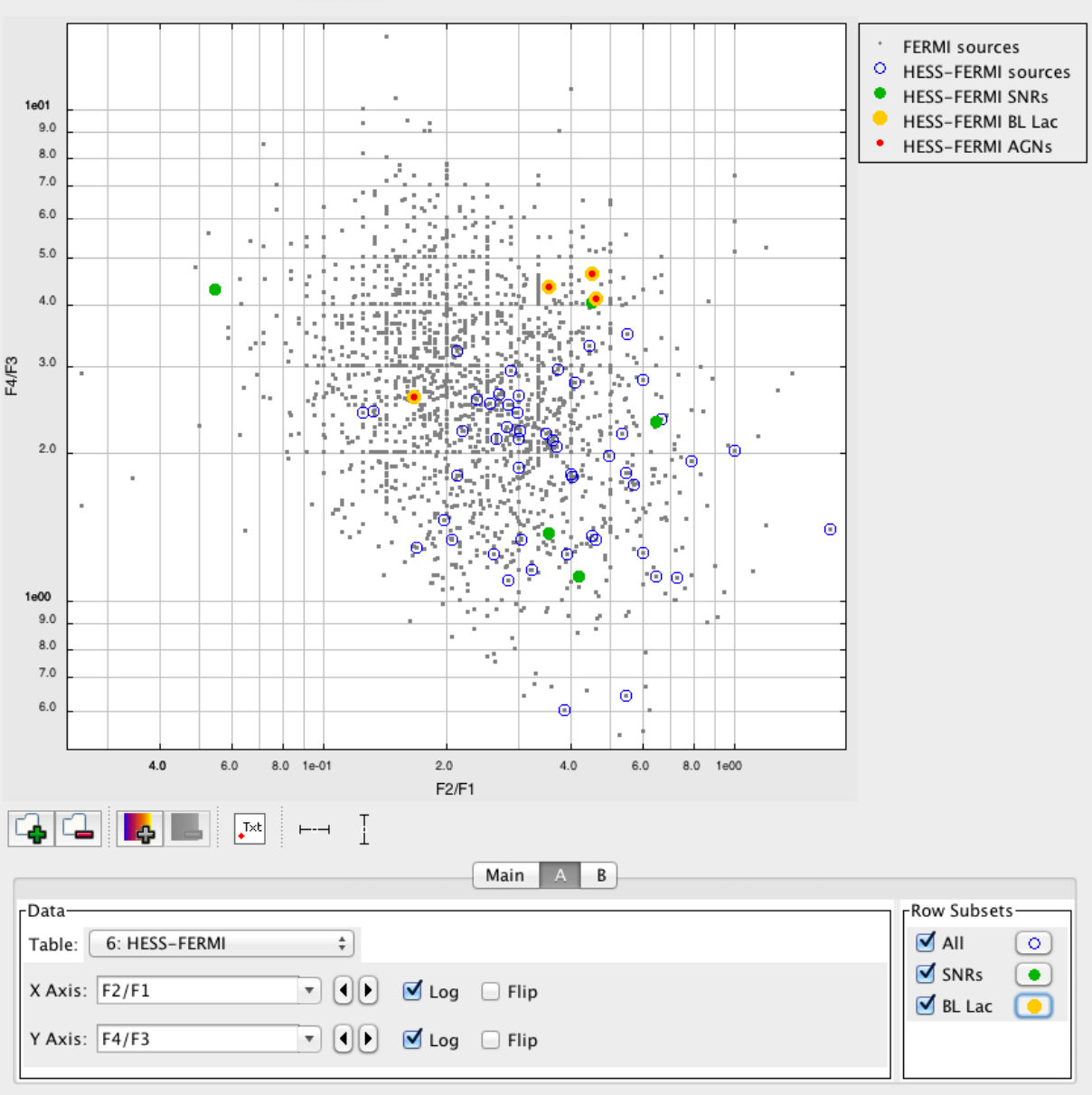}
  \caption{Left: overlay of the H.E.S.S. source catalog and the second Fermi-LAT catalog in Aladin. Right: Cross-matching between the H.E.S.S. and Fermi-LAT catalogs and identification of individual source classes in a ``high-energy color-color plot'' in TOPCAT.}
  \label{fig:hess-lat-image}
\end{figure*}

We start by investigating the H.E.S.S. source catalog looking for counterparts of the H.E.S.S. objects in the second Fermi-LAT catalog. The H.E.S.S. catalog is queried using the SIMBAD portal and the result is saved in a {\it VOTable}. The table can then be loaded into the Aladin tool \citep{CDS2011,Boch2011} dedicated to the analysis of astronomical images. In Aladin it is possible to project the list of H.E.S.S. sources onto the Fermi-LAT sky (see Fig.~\ref{fig:hess-lat-image}). 

\begin{figure*}
  \centering
  \includegraphics[width=7cm]{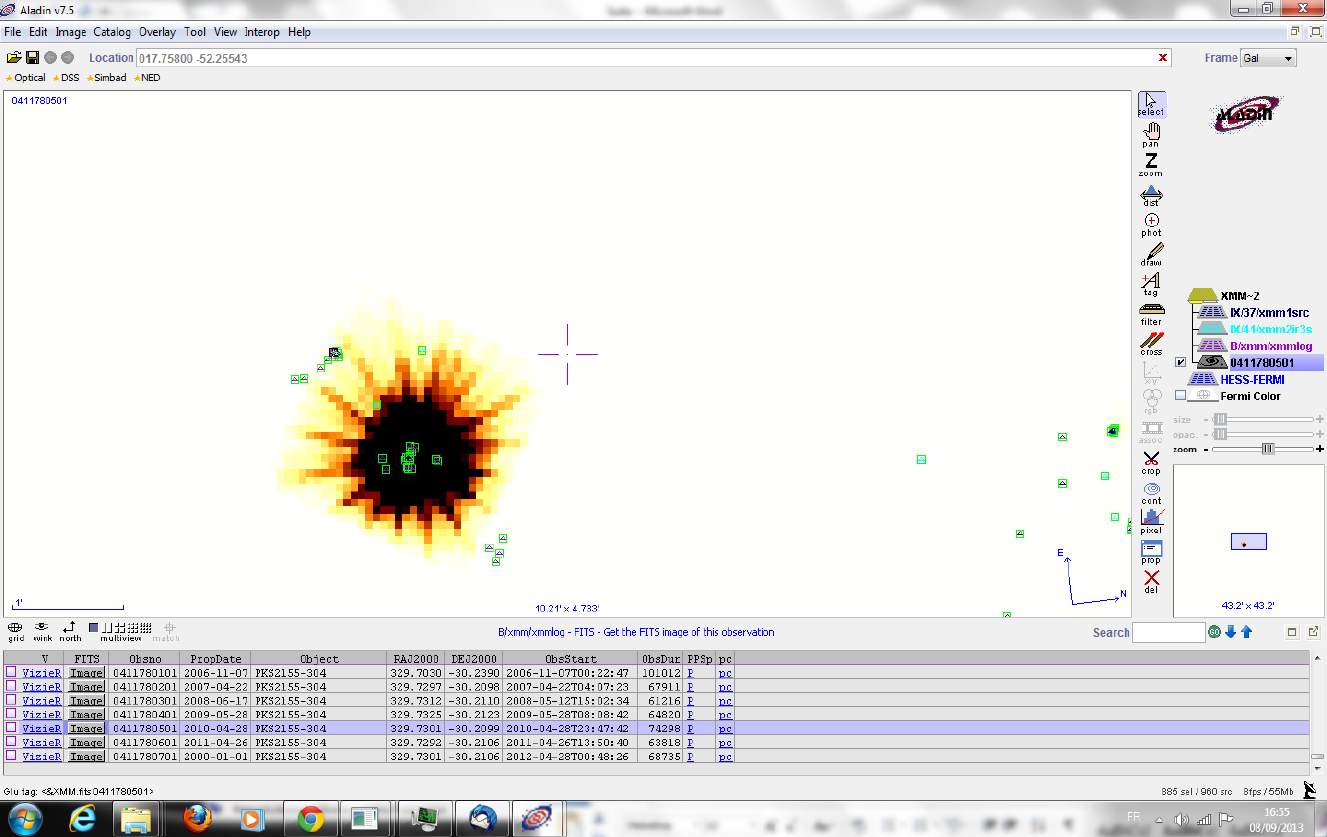}
  \includegraphics[width=9cm]{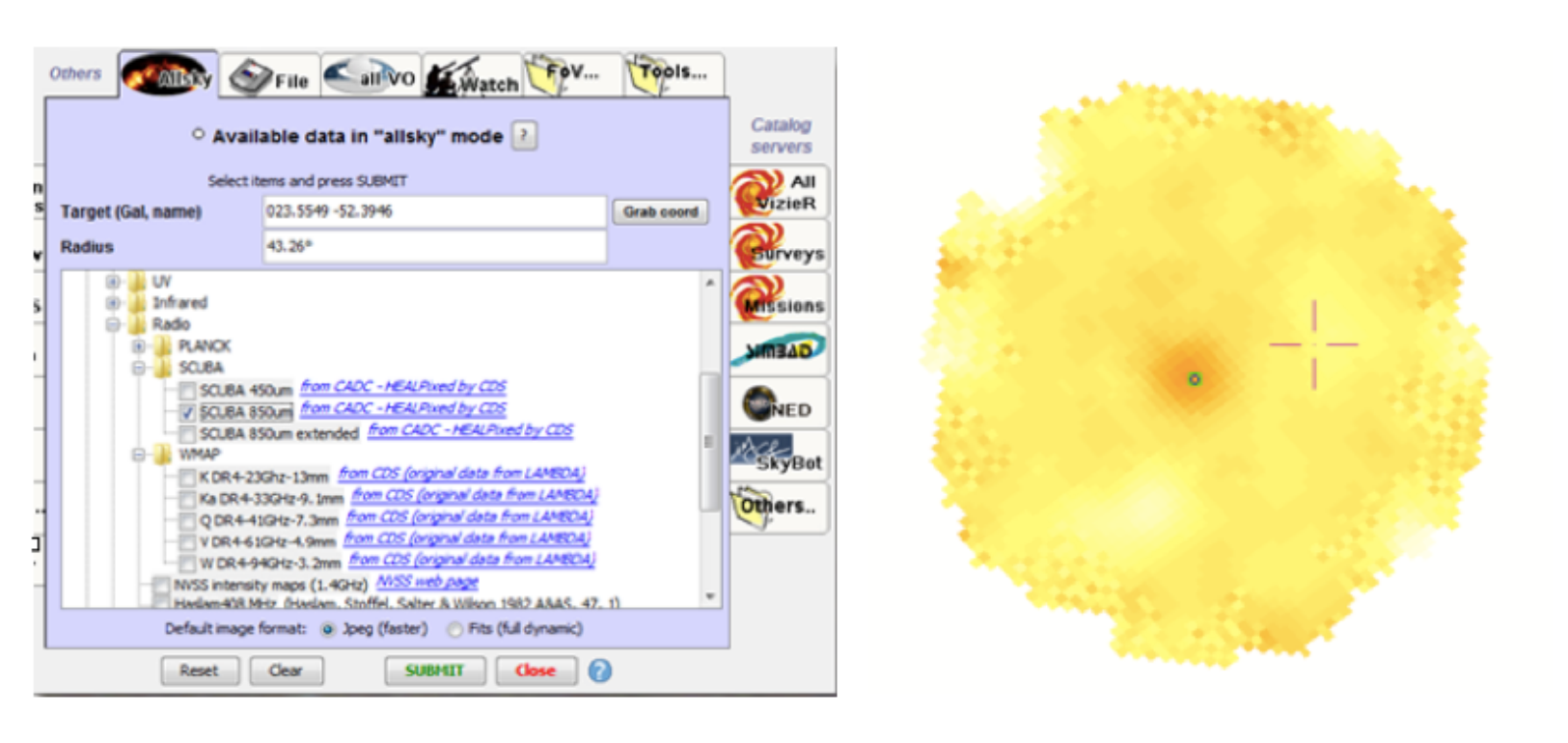}\\
  \includegraphics[width=12cm]{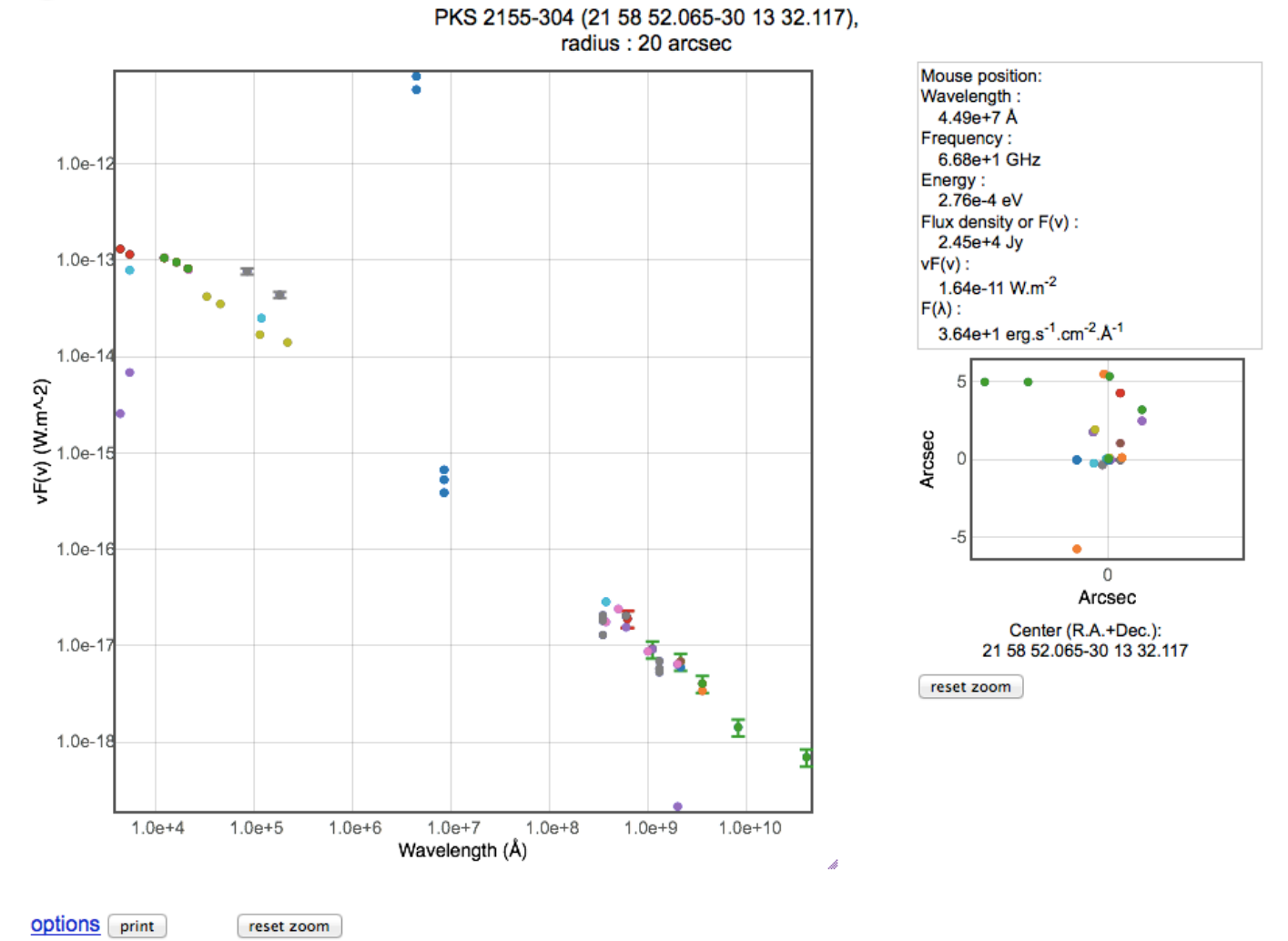}
  \caption{Top left: X-ray image (0.2-12~keV) of PKS~2155-304 recovered from the XMM-Newton archive in Aladin. Top right: Analogous image from SCUBA in the sub-mm range (850 microns). Bottom: broad-band photometry data for PKS~2155-304 assembled using the VizieR photometry viewer widget.}
  \label{fig:PKS-XMM-SCUBA}
\end{figure*}

The 1873 Fermi-LAT sources are loaded in TOPCAT \citep{Taylor2011} by querying the VizieR catalogue service \citep{LO2012}. We broadcast the H.E.S.S. sources from Aladin to TOPCAT using the SAMP protocol \citep{SAMP}. 
Next, we want to cross match both catalogs and possibly identify catalog entries describing the same source, at least when projected onto the plane of the sky. A difference in radial distance may exist but is unlikely given the relatively sparse distribution of H.E.S.S. and Fermi-LAT sources across the sky. The cross matching is carried out inside the VO tool TOPCAT assuming that a H.E.S.S. source and its supposed Fermi-LAT counterpart have an apparent angular distance of less then 1$^\circ$. The cross matching establishes a new catalog that at the time of this writing contains 61 combined H.E.S.S. and Fermi-LAT sources. In the following we name this catalog {\it HESS-Fermi}.

The HESS-Fermi catalog still contains various types of sources. We want to classify them according to their overall gamma ray properties. In stellar astronomy, a convenient way to classify stars is to plot their colors against each other. Different stellar types then occur at defined regions of such a color-color diagram. We follow the equivalent approach for our HESS-Fermi sources by plotting different gamma ray flux ratios against each other. The ``color-color'' diagram to be computed involves the Fermi-LAT fluxes integrated over the following energy bands: 100 -- 300 MeV (F1), 300 -- 1000 MeV (F2), 1 -- 3 GeV (F3), 3 -- 10 GeV (F4). After constructing a plot of F4/F3 as a function of F2/F1 in log--log scale, the HESS table entries allow us to associate a type to each 
source (see Fig.~\ref{fig:hess-lat-image}, right). It turns out that the sources detected by both H.E.S.S. and Fermi-LAT preferentially populate the bottom-right corner of the color-color diagram. For now, the number of identified sources of a given type is too low to make a clear statement about its preferred locus on the plot.

\cite{Ackermann2011} have produced a catalogue of active galactic nuclei (AGNs), and we can identify sources from our HESS-Fermi catalog that are AGNs by 
matching their Fermi-LAT identifiers in TOPCAT. This yields 4 known AGNs in our sample, highlighted in the right panel of Fig.~\ref{fig:hess-lat-image}.

\section{Constructing multi-wavelength images and spectra for the blazar PKS~2155-304}

The remainder of the tutorial focuses on a particular object of the self-constructed Fermi-HESS catalog. PKS~2155-304 is a BL Lac object that is variable across a wide spectral range \citep[see][and references therein]{Abramowski2014}. BL Lac objects are AGNs launching magnetized, ballistic jets that are nearly aligned with the line of sight of the observer. The jet emission can be modeled as relativistically beamed synchrotron-self-Compton emission: the very hot electrons produce synchrotron emission that peaks in the radio band and then the very same electrons Comptonize the synchrotron photons and up-scatter them to gamma-rays and beyond.

The tutorial shows how to obtain multi-waveband imaging for PKS~2155-304 starting from the displayed catalog HESS-Fermi in Aladin. The interoperability of the VO tools allows the user to recover in a straightforward manner X-ray imaging data from XMM-Newton \citep{Lumb2012} and a sub-mm image from SCUBA \citep{Holland1998} by querying the respective data archives. The resulting images are given in Fig.~\ref{fig:PKS-XMM-SCUBA} (top panel). 

Aside from imaging at multiple wavebands, a photometry widget developed for the CDS portal allows to retrieve at once broad-band photometry data from many VizieR catalogues for a given object. Applying it on PKS~2155-304 we obtain a photometric ``spectrum'' of the source that reaches from the radio band all the way to the gamma ray range (see Fig.~\ref{fig:PKS-XMM-SCUBA}, bottom).

\section{Summary and concluding remarks}

We have briefly presented a tutorial illustrating the use of several VO tools in the context of gamma ray astronomy. The VO tools allow the user to efficiently mine data bases for images and spectra across all observed wave bands. The VO can thus be of great help when studying large ensembles of objects as well as for the in-depth exploration of an individual source.

The development of the VO tools is guided by the needs of the community. The Strasbourg Data Centre therefore is very grateful for feed-back and suggestions to improve and further develop the VO services.





%
\end{document}